# Apodizer Design To Efficiently Couple Light Into A Fiber Bragg Grating

Jimmie Adriazola [*] Roy H. Goodman [†]

January 2, 2023

**Abstract.** We provide an optimal control framework for efficiently coupling light in a bare fiber into Bragg gratings with a cubic nonlinearity. The light-grating interaction excites gap solitons, a type of localized nonlinear coherent state which propagates with a central frequency in the forbidden band gap, resulting in a dramatically slower group velocity. Due to the nature of the band gap, a substantial amount of light is back-reflected by the grating's strong reflective properties. We optimize, via a projected gradient descent method, the transmission efficiency of previously designed nonuniform grating structures in order to couple more slow light into the grating. We further explore the space of possible grating designs, using genetic algorithms, along with a previously unexplored design parameter: the grating chirp. Through these methods, we find structures that couple a greater fraction of light into the grating with the added bonus of creating slower pulses.

**1. Experimental and Technological Context.** Dramatic slowing of light has been observed in a wide variety of experimental settings over the past two decades. This phenomenon offers enticing technological applications including efficient optical switches, sensitive interferometry, and optical quantum memory [12]. Of the several experimental platforms which can generate slow light, fiber Bragg gratings (FBGs) offer the considerable advantage of having structural properties which can be tailored specifically to the characteristics of an incoming light source.

Modern optical communications systems already use FBGs as notch filters and as components in optical add-drop multiplexers [6]. Optical fibers transmit information as coherent pulses of light that must be manipulated or redirected as they travel, and switches which make use of slow light may produce faster data transmission with lower power requirements [20]. Additionally, more orderly flow of data transmission in optical networks can be achieved by using slow light to control delay times [37]. Therefore, FBG technology that can significantly slow down, or even halt, light is highly desirable.

An FBG is an optical element whose index of refraction varies periodically, see Figure 1.1. The grating enables the strong dispersion of light over a short distance due to a resonance between the grating's period $\Lambda$ and electromagnetic wavelengths near the Bragg wavelength $\lambda_B = 2\Lambda$. In so-called chirped gratings, discussed in further detail in Section 2.1 and shown in Figure 1.1, the period $\Lambda$ is spatially varying.

The effect of an FBG is to strongly couple forward and backward propagating waves near the resonant wavelength [11]. This creates a photonic bandgap, i.e., an interval of frequencies at which low-amplitude light is reflected and unable to propagate. This bandgap is centered at the Bragg angular frequency $\omega_B = \frac{\pi c}{\bar{n}\Lambda}$, where $c$ is the speed of light in vacuum and $\bar{n}$ is the mean index of refraction.

---

[*]Department of Mathematical Sciences, New Jersey Institute of Technology (current affiliation: Department of Mathematics, University of California, Santa Barbara) corresponding email: jadriazola@ucsb.edu

[†]Department of Mathematical Sciences, New Jersey Institute of Technology

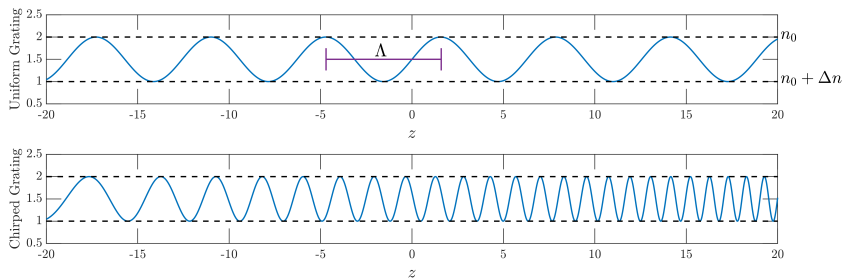

Fig. 1.1. *Top: Schematic showing a spatially varying index of refraction $n(z)$ that periodically varies between a reference index $n_0 = 1$ to $n_0 + \Delta n = 2$. Bottom: Schematic of a chirped grating, i.e., a grating with a spatially dependent period $\Lambda(z)$.*

In materials with a Kerr nonlinearity, the index of refraction responds proportionally to the electric field intensity [25]. When the nonlinearity is large enough and acts over long enough times and distances to balance with dispersion, high intensity light shifts the Bragg frequency [11]. For positive nonlinearities, the refractive index increases with intensity, thereby shifting the bandgap downward. Therefore, at the edges of a high intensity region of light, the light is continuously Bragg reflected into the high intensity region; see Figure 1.2 for a schematic diagram. Light then propagates through the grating seemingly unimpeded. Systems, such as these, that exhibit coherent structures arising from a balance between nonlinear effects and dispersion typically support solitary waves.

Indeed, Aceves and Wabnitz constructed a two-parameter class of solitary wave solutions [1], often called Bragg solitons to distinguish them from the classical notion of a soliton [26]. These waves, experimentally discovered by Eggleton, et al., in 1996 [13, 14], are modeled as solutions to evolution equations derived from Maxwell's equations using coupled-mode theory, which we briefly discuss in Section 2, and can travel with a speed anywhere from the speed of light in the medium down to zero. The existence of Bragg solitons demonstrates the possibility of slow light in an FBG. While Bragg solitons can in theory propagate at slow speeds, it is difficult to initialize such waves experimentally: to create a Bragg soliton, one must input light at a frequency inside the bandgap, and such frequencies are strongly reflected. Neglecting nonlinear effects, the FBG essentially acts as a band-stop filter, reflecting wavelengths whose frequency is within the bandgap.

To overcome this, Mok, et al. [24], building on previous work by Lenz, et al. [21], use a two-pronged strategy to couple light into an FBG. First, they use an apodized grating, i.e., the grating strength is ramped up gradually from zero. Secondly, they input so-called out-gap solitons, wave packets with a mean frequency outside of the bandgap. As a result of the apodization, the light coupled into the grating has its frequency gradually shifted into the bandgap. Although this experiment is the first of its type and generated a pulse with a group velocity 16% that of light in glass, this pulse contained only about 20-30% of the the input energy, while the remaining light was reflected, rendering the setup highly inefficient.

In order to address this inefficiency, Rosenthal and Horowitz [29], designed a two-segment apodization function that allowed the creation of a pulse that retains about 68% of the incident energy and with a speed roughly 3.2% of light speed. Despite this remarkable improvement in efficiency, the authors provide limited mathematical detail about the process behind discovering such a design. This leaves room for quantitative

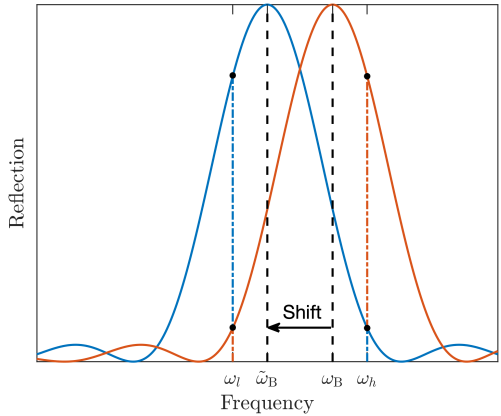

FIG. 1.2. *A schematic of the intensity-dependent bandgap shift and its effect on the reflectivity of light. Higher intensities decrease the Bragg frequency. Shown is an example shift from some initial frequency $\omega_B$ to another denoted by $\tilde{\omega}_B$. Light at the lower frequency $\omega_l$ is initially more strongly Bragg reflected. Meanwhile, light at the higher frequency $\omega_h$ experiences a smaller reflection.*

investigations into the efficiency of generating slow light using FBGs.

In this article, we improve on the method of Rosenthal and Horowitz by mathematically formulating an appropriate optimal control problem whose objective is to design an apodization profile that maximizes the coupling of light into the FBG. By including a spatially varying chirp profile in the optimization problem, an element to the design considered by Rosenthal and Horowitz in an earlier work [28], we achieve a higher coupling efficiency of 82.6% transmission, while further reducing the pulse speed to about 0.5% the speed of light in glass. Although the gratings reported in this manuscript would require a higher accuracy in their fabrication in a laboratory setting, work due to Sahin, et al. [30,31] demonstrates that such fabrication is possible.

This work is organized as follows: in Section 2, we provide the physical model and give precise details of past numerical experiments we aim to improve upon. We attempt to gain intuition behind these simulations by fitting the numerical data to the Aceves and Wabnitz waveform. In Section 3, we formulate the optimal control problem which seeks to maximize the energy transmitted into the fiber while treating the grating structure as the control. We provide the necessary optimality conditions for the control problem, use these conditions to design grating structures using numerical methods discussed in Appendix A, and present numerical results in Section 4 while discussing directions for future study in Section 5.

**2. The Physical and Numerical Model.**

**2.1. Brief Overview of Coupled-Mode Theory.** The evolution of an electric field propagating in an optical fiber can be effectively modeled by the one dimensional nonlinear wave equation

$$\partial_\tau^2 \left( n^2(z, |E|^2) E \right) = \partial_z^2 E, \tag{2.1}$$

in dimensionless units where the speed of light $c = 1$, and where $z$ denotes the axial direction of propagation. In the context of Figure 1.1, we choose the reference index of refraction $n_0 = 1$. Let $\varepsilon$ be a small contrast of the index of refraction $n$ so that it can be modeled, following [16], as

$$n^2 = 1 + \varepsilon \left( \nu(\varepsilon z) \cos(2k_B z + 2\Phi(\varepsilon z)) + |E|^2 \right). \tag{2.2}$$

Here the coefficient $\nu$ describes the strength of the grating and $\Phi'$ describes the chirp, i.e., the local modulation of the grating's wavelength. The final term describes a Kerr nonlinearity with small Kerr coefficient $\varepsilon$.

Using the method of multiple scales with the ansatz,

$$E = \sqrt{\varepsilon}\left(u(\varepsilon z, \varepsilon\tau)e^{i(k_B(z-\tau)+\Phi)} + v(\varepsilon z, \varepsilon\tau)e^{-i(k_B(z+\tau)+\Phi)}\right) + \mathcal{O}\left(\varepsilon^{\frac{3}{2}}\right) \quad (2.3)$$

and letting $x = \varepsilon z$, $t = \varepsilon\tau$, denote the slow variables, the following system of hyperbolic equations, known as the nonlinear coupled-mode equations (NLCME),

$$\begin{aligned} i\partial_t u + i\partial_x u + \kappa(x)v + \eta(x)u + \left(|u|^2 + 2|v|^2\right)u &= 0, \\ i\partial_t v - i\partial_x v + \kappa(x)u + \eta(x)v + \left(2|u|^2 + |v|^2\right)v &= 0. \end{aligned} \quad (2.4)$$

arise as solvability conditions on the forward and backward slowly varying envelopes $u(x,t)$ and $v(x,t)$, respectively [16]. The coefficient $\kappa(x)$ is proportional to the local strength $\nu(x)$ of the grating while $\eta(x)$ is proportional to the local chirp $\Phi'(x)$. We refer to regions where $\kappa(x)$ and $\eta(x)$ vanish as the bare fiber to indicate the absence of the grating.

For low-amplitude light, the NLCME reduce to a set of linear coupled-mode equations

$$\begin{aligned} i\partial_t u + i\partial_x u + \kappa(x)v + \eta(x)u &= 0, \\ i\partial_t v - i\partial_x v + \kappa(x)u + \eta(x)v &= 0. \end{aligned} \quad (2.5)$$

When $\kappa$ and $\eta$ are constant, this system has a dispersion relation given by

$$\Omega(Q) = \eta \pm \sqrt{Q^2 + \kappa^2}. \quad (2.6)$$

Introducing a chirp $\eta(x)$ into the grating shifts the center of the bandgap, i.e., the set of frequencies $\Omega \in \left(\eta - \sqrt{Q^2 + \kappa^2}, \eta + \sqrt{Q^2 + \kappa^2}\right)$ which, for fixed $\kappa$ and $\eta$, do not satisfy the dispersion relation for any wavenumber $Q$. We show an example of dispersion relation (2.6) in Figure 2.1.

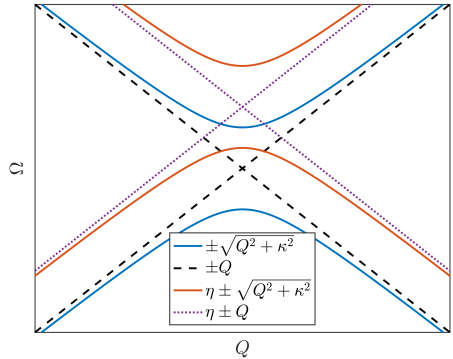

FIG. 2.1. *The dispersion relation (2.6) of the linear coupled-mode equations (2.5) both with and without a chirp, demonstrating how the chirp shifts the bandgap.*

In the case of a uniform grating, that is, where $\kappa \equiv \kappa_0$ and $\eta \equiv 0$, the NLCME admit a two-parameter family of translationally invariant solitary wave solutions, called Bragg solitons [1],

$$u_B = \sqrt{\frac{\kappa_0(1+c)}{3-c^2}} \left(1-c^2\right)^{1/4} W(X)\exp\left(i\phi(X) - iS\cos\theta\right),$$
$$v_B = -\sqrt{\frac{\kappa_0(1-c)}{3-c^2}} \left(1-c^2\right)^{1/4} W^*(X)\exp\left(i\phi(X) - iS\cos\theta\right),$$
(2.7)

where

$$X = \kappa_0 \left(1-c^2\right)^{-1/2} (x - ct),$$
$$S = \kappa_0 \left(1-c^2\right)^{-1/2} (t - cx),$$
(2.8)
$$\phi(X) = \frac{4c}{3-c^2}\arctan\left(\tanh(X\sin\theta)\tan\frac{\theta}{2}\right),$$
$$W(X) = \sin\theta\,\mathrm{sech}\left(X\sin\theta - \frac{i\theta}{2}\right),$$

with free parameters $0 \le \theta \le \pi$ and $-1 < c < 1$. The dependence of the Bragg soliton on the parameters is quite complicated, but we can make a few observations. The parameter $c$ describes the velocity of the pulse, appears in a Lorentz contraction, and, through the factors $(1\pm c)^{1/2}$, controls the relative amplitude of the forward and backward envelopes. Note that for stationary Bragg solitons, i.e., when $c = 0$, the frequency of the stationary oscillation is given by $\kappa_0 \cos\theta$ so as $\theta$ is increased from 0 to $\pi$, the frequency of the standing wave moves from the right edge of the band gap to the left edge, while always remaining inside the gap.

Because we are considering the case where $\kappa(x)$ and $\eta(x)$ vary in space, it is reasonable to ask if the linear operator in Equation (2.5) supports any localized eigenfunctions. Such discrete spectrum is discussed in Refs. [16,17] and requires that $\kappa(x)$ approaches nonzero values as $x \to \pm\infty$, so that the spectrum has a gap. Because in this paper we consider gratings whose amplitude vanishes as $x \to -\infty$, there is no gap, and thus no mechanism to support localized eigenfunctions.

The constant coefficient NLCME possess two conserved quantities, an energy

$$E = \int_{-\infty}^{\infty} \mathcal{E}\,dx := \int_{-\infty}^{\infty} \left(|u|^2 + |v|^2\right) dx,$$
(2.9)

and a momentum

$$P = \int_{-\infty}^{\infty} \mathcal{P}\,dx := i\int_{-\infty}^{\infty} \left(u\partial_x u^\dagger + v\partial_x v^\dagger\right) dx,$$
(2.10)

where $\dagger$ denotes complex conjugation. Allowing the coefficients to vary in space breaks the translation invariance, so that conservation of momentum fails to hold, yet energy conservation remains. Because of this fact, we make use of the energy (2.9) in posing an optimal control problem in Section 3.

**2.2. Previous numerical experiments and their reinterpretation.** In this section we discuss two earlier studies of light propagation into an apodized FBG, namely those of Mok et al. and of Rosenthal and Horowitz [24, 29]. Both groups numerically simulate solutions to Equations (2.4) for particular grating profiles $\kappa(x)$ and without chirp. We reproduce those simulations here, and provide some additional post-processing in order to understand them better.

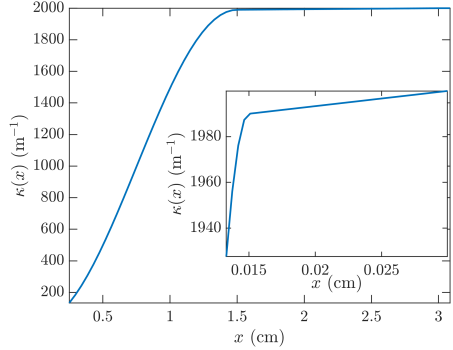

Fig. 2.2. *The Rosenthal and Horowitz apodization function* (2.11), *with* $\zeta = 0.995$, $\kappa_0 = 2\text{mm}^{-1}$, *and* $L_1 = L_2 = 1.5\text{cm}$.

In these simulations, the profile $\kappa(x)$ grows from $\kappa = 0$ for $x \leq 0$ to a value $\kappa_0$ for $x \geq a > 0$. The pulse is launched with a fixed profile from a point $x = x_\text{input} < 0$ in the bare fiber and has a momentum directed toward the apodization interval $0 \leq x \leq a$. The pulse is launched with a fixed profile from a point $x = x_\text{input} < 0$ in the bare fiber and has a momentum directed toward the apodization interval $0 \leq x \leq a$. These simulations do not include a chirp $\eta(x)$, so we set $\eta \equiv 0$ in this section and postpone discussions of the chirp $\eta$'s role until we define the relevant optimal control problem in Section 3.

Both previous studies use an apodization profile of the form

$$(2.11) \qquad \kappa(x) = \begin{cases} \frac{\zeta \kappa_0}{2}\left(1 - \cos\frac{\pi x}{L_1}\right) & 0 < x \leq L_1 \\ \zeta \kappa_0 + \frac{\kappa_0}{L_2}(1-\zeta)(x - L_1) & L_1 < x < a = L_1 + L_2, \end{cases}$$

with one key difference. In Ref. [24], $\zeta = 1$, so that the apodization takes place entirely on the interval $[0, L_1]$. Such an apodization profile is known as a raised cosine. By contrast, the simulations of Ref. [29] use $\zeta = 0.995$, a very small modification, that a shallow linear ramp following the raised cosine.

The numerical simulation in [29] takes the form of a signaling problem, i.e., the solution is initialized by a time-dependent boundary condition at the "input" endpoint. The simulation also assumes an initial condition that is identically zero outside of the input. The signaling data is given by the profile

$$(2.12) \qquad u(x_\text{input}, t) = A\,\text{sech}\left(\frac{t-\phi}{\sigma}\right) e^{-i\Omega t}, \quad v(x, 0) = 0,$$

where $x_\text{input}$ denotes the left-most endpoint of the spatial domain. It propagates through an initial segment of fiber with no grating with an oscillatory frequency outside the band gap of the dispersion relation (2.6) caused by the grating of amplitude $\kappa_0$.

Before showing the profile and the numerical simulations, it is important that we reinterpret the nondimensional form of NLCME (2.4) with physically realistic dimensions. The choices of the simulation parameters, in SI units, are $L_1 = 1.5\text{cm}$, $L_2 = 1.5\text{cm}$, $\kappa_0 = 2\text{mm}^{-1}$, while the signalling data parameters are $A = 16.4\text{W}$, $\sigma = 96.4\text{ps}$, $\phi = 4\text{ns}$, and $\Omega = 0.398\text{GHz}$.

We perform these simulations using a second-order-in-time operator splitting method, detailed in Appendix A.3, to solve Equation (2.4) with a spatial discretization of 4000 points and temporal discretization of 12000 points. We solve Equation (2.4) out to 6 ns, set $x_{\text{input}} = -20$ cm, and set the right endpoint to 16 cm. The numerical method and majority of these parameters will remain consistent throughout this work, unless otherwise noted.

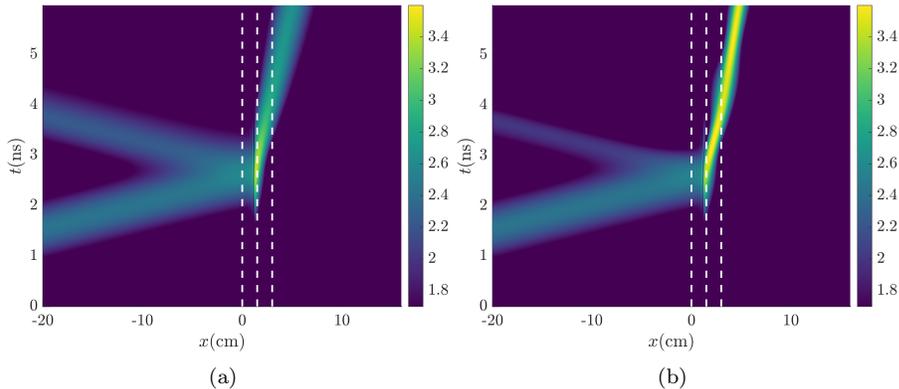

Fig. 2.3. *Numerical simulations of Equation* (2.4), *shown on a logarithmic scale, with the apodization design* (2.11) *consistent with the parameters detailed in the text. Dashed lines provided to help visualize the regions in space over which the two segment apodization varies, cf.* (2.11) *and Figure 2.2. Panel (a) corresponds to the original Mok, et al., design, i.e.,* $\zeta = 1$, *while Panel (b) has* $\zeta = 0.995$.

Figure 2.3 shows our reproduction of both Mok's and Rosenthal's numerical simulations. It demonstrates the vast improvement in energy transmission that results from the choice $\zeta = 0.995$, instead of $\zeta = 1$, in the apodization function (2.11). It is clear from these figures that the fraction of transmitted light increases between the two simulations while the fraction of reflected light decreases. Specifically, the total energy transmission is increased from about 20% in the former case to about 66% in the latter.

Rosenthal and Horowitz explain the remarkable performance of the two-segment apodization function by appealing to ideas from soliton perturbation theory. They argue that the second segment of the apodization function *adiabatically* shifts the high-intensity waveform, initially at $x = L_1$, into the band gap, and, as a result, minimizes back-reflection. We believe the ad hoc nature of this argument invites further study into such Bragg grating designs in two directions. To make the adiabatic claim precise and provide further physical insights is one such direction and should be pursued in future work. This paper is focused on another direction; to more thoroughly investigate the space of grating structures through a systematic optimization.

To interpret the result shown in Figure 2.3(b), we fit spatial solitary waves of the form (2.7) to the numerical simulation data at specific times and display results in Figure 2.4. These fits do not show particularly convincing evidence that the excited waveforms are Bragg solitons, especially when compared with the analogous figures in Section 4 showing results from an optimization.

To demonstrate the need for using optimal control to further improve transmission efficiency, we first perform a brute force optimization over the parameters in the family of apodization profiles given by Equation (2.11). In particular, we define a parameter $\xi \in (0, 1)$, so that

$$(2.13) \qquad L_1 = 3(1-\xi) \text{ cm}, \quad L_2 = 3\xi \text{ cm},$$

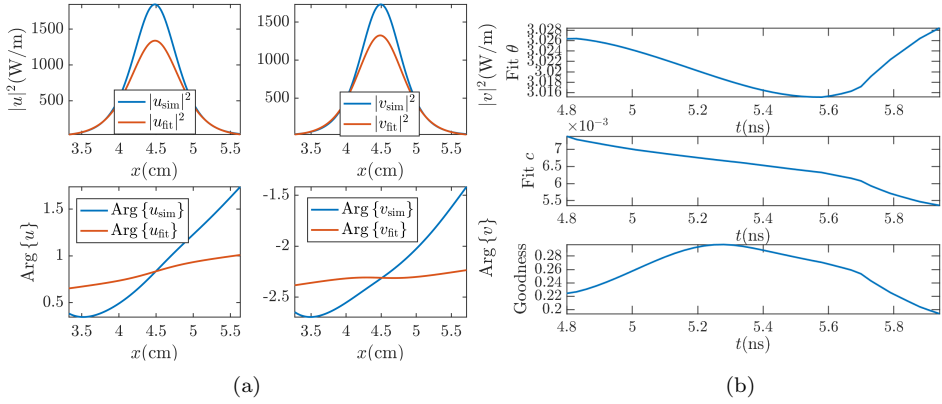

(a)  (b)

FIG. 2.4. *A least-squares fit of a Bragg soliton of the form* (2.7) *to the Rosenthal and Horowtiz simulation at the instant the pulse exits the grating at t = 4.8 ns, as shown in Figure* 2.3. **(a)** *The power density and local phase of the coupled-modes u and v.* **(b)** *The parameters defining the Bragg soliton in Equation* (2.7), *showing a near constant value in the phase parameter θ and a slow speed parameter c. The goodness-of-fit over time is also shown.*

and optimize the profile over the parameters $(\xi, \zeta)$. Note that the Rosenthal and Horowitz parameters are $(\xi, \zeta)_{\text{RH}} = (0.5, 0.995)$ in Figure 2.3. A simple minimization finds the optimal parameters $(\xi, \zeta)_* = (0.567, .99244)$, which yield an improved transmission near 68%. We interpret this as an indication that optimal control theory will be able to further improve the transmission efficiency.

### 3. Optimal Control Formulation.

**3.1. Objective.** We now precisely formulate an optimal control problem whose objective is to find the grating structures that maximize the transmission of light into the constant grating portion of the optical fiber. To this end, we make use of the following local conservation law

$$(3.1) \qquad \partial_t \mathcal{E} + \partial_x \mathcal{F}_\mathcal{E} = 0,$$

where $\mathcal{F}_\mathcal{E}$ is the local energy flux. Note that this conservation law is the differential form of energy conservation, where the energy is given by Equation (2.9).

We treat the grating functions $\kappa(x)$ and $\eta(x)$ as the control functions, and assume that the apodization region is of fixed width $a > 0$, consistent with the Rosenthal and Horowitz apodization function (2.11). In addition, we use parameters consistent with those of Figure 2.2, including the signalling data (2.12), length of the spatial domain, duration of simulation time, and number of discretization points. We search over the admissible class $\mathcal{C}$ of grating functions which we define as the space of absolutely continuous functions such that

$$(3.2) \qquad \kappa(x) = \begin{cases} 0 & x \leq 0, \\ \kappa_0 & a \leq x, \end{cases} \quad \text{and} \quad \eta(x) = \begin{cases} 0 & x \leq 0, \\ 0 & a \leq x. \end{cases}$$

Now, the optimal control problem we seek to solve is

$$(3.3) \qquad \min_{(\kappa,\eta) \in \mathcal{C}} \mathcal{J} = \min_{(\kappa,\eta) \in \mathcal{C}} \left\{ -\int_0^T \mathcal{F}_\mathcal{E}(u,v;a)dt + \frac{\gamma}{2} \int_0^a \left( (\partial_x \kappa)^2 + (\partial_x \eta)^2 \right) dx \right\},$$

subject to the differential equation constraint (2.4). Note that the time horizon $T$ is chosen ad hoc. We found that $T = 6$ ns strikes a good balance between a time allowance for solitons to penetrate into the constant grating portion of the fiber, in each of the simulations that will follow, yet is a time small enough to keep simulations from becoming too computationally intensive.

Although the first term in the objective $\mathcal{J}$ is a cost that runs over time, the following simple calculation demonstrates that this term can alternatively be written as

$$(3.4) \quad \int_0^T \mathcal{F}_\mathcal{E}(u,v;a)dt = \int_0^T \int_\infty^a \partial_x \mathcal{F}_\mathcal{E} dx dt = \int_a^\infty \int_0^T \partial_t \mathcal{E} dt dx = \int_a^\infty \mathcal{E}(x,T)dx,$$

by the fundamental theorem of calculus, Fubini's theorem, and conservation law (3.1). In this sense, the term which promotes a greater energy flux into the constant grating portion can be written as a running cost of terminal energy in space. The second term in objective $\mathcal{J}$ is called a Tikhonoff regularization and is taken over space. Such a regularization term is used extensively in studies of ill-conditioned optimal control and inverse problems [2, 18, 34]. Its effect is to penalize rapid variations in the grating which would be infeasible to design, in practice.

**3.2. Necessary Optimality Conditions.** In order to solve Problem (3.3), we use a line search discussed in Appendix A.2. Part of the method involves the computation of the gradient which depends on a suitably defined criterion for optimality. To facilitate such a computation, let $H(x-a)$ denote Heaviside's function. Following the calculation shown in Equation (3.4), we find

$$(3.5) \quad \int_a^\infty \mathcal{E}(x,T)dx = \int_{-\infty}^\infty H(x-a) \int_0^T \partial_t \mathcal{E} dt dx$$
$$= \int_{-\infty}^\infty H(x-a) \int_0^T \left( u^\dagger \partial_t u + u \partial_t u^\dagger + v^\dagger \partial_t v + v \partial_t v^\dagger \right) dt dx.$$

Now, define the Lagrangian density by

(3.6)
$$\mathcal{L} = \operatorname{Re}\left\langle \lambda, i\partial_t u + i\partial_x u + \kappa(x)v + \eta(x)u + \left(|u|^2 + 2|v|^2\right)u \right\rangle_{L^2([0,T])}$$
$$+ \operatorname{Re}\left\langle \mu, i\partial_t v - i\partial_x v + \kappa(x)u + \eta(x)v + \left(2|u|^2 + |v|^2\right)v \right\rangle_{L^2([0,T])}$$
$$- H(x-a)\int_0^T \left( u^\dagger \partial_t u + u\partial_t u^\dagger + v^\dagger \partial_t v + v\partial_t v^\dagger \right) dt + \frac{\gamma}{2}I(x;a)\left((\partial_x \kappa)^2 + (\partial_x \eta)^2\right),$$

where $\lambda$ and $\mu$ are Lagrange multipliers and $I(x;a) = H(x) - H(x-a)$. The objective in optimal control problem (3.3) is then written in the form

$$(3.7) \quad \mathcal{J} = \int_\mathbb{R} \operatorname{Re}\left\{ \mathcal{L}\left(u, v, \partial_t u, \partial_t u^\dagger, \partial_t v, \partial_t v^\dagger, \partial_x u, \partial_x v, \kappa, \eta, \partial_x \kappa, \partial_x \eta, \lambda^\dagger, \mu^\dagger\right) \right\} dx$$

so that the optimization is expressed as a saddle point problem which enforces $u$ and $v$ solve NLCME (2.4). Using typical arguments from the classical calculus of variations [15], we find the desired optimality conditions by taking the appropriate functional derivatives. Setting functional derivatives with respect to the state variables to zero gives

$$(3.8) \quad \frac{\delta \mathcal{J}}{\delta u} + \frac{\delta \mathcal{J}}{\delta u^\dagger} = 0, \quad \frac{\delta \mathcal{J}}{\delta v} + \frac{\delta \mathcal{J}}{\delta v^\dagger} = 0.$$

These conditions imply that $\lambda$ and $\mu$ satisfy the following linear PDE

(3.9a) $\quad i\partial_t\lambda + i\partial_x\lambda + \left(\eta + 2\mathcal{E} + u^{\dagger 2}\right)\lambda + \left(\kappa + 4v^\dagger\operatorname{Re}\{u\}\right)\mu = 0,$

(3.9b) $\quad i\partial_t\mu - i\partial_x\mu + \left(\eta + 2\mathcal{E} + v^{\dagger 2}\right)\mu + \left(\kappa + 4u^\dagger\operatorname{Re}\{v\}\right)\lambda = 0.$

An integration by parts in the derivation of these Euler-Lagrange equations yields boundary terms which must also be set to zero:

(3.10) $\quad \left.(\partial_{\partial_t u}\mathcal{L} + \partial_{\partial_t u^\dagger}\mathcal{L})\right|_{t=T} = 0, \quad \left.(\partial_{\partial_t v}\mathcal{L} + \partial_{\partial_t v^\dagger}\mathcal{L})\right|_{t=T} = 0,$

directly implying that

(3.11a) $\quad \lambda(x,T) = 2iH(x-a)\operatorname{Re}\{u(x,T)\},$

(3.11b) $\quad \mu(x,T) = 2iH(x-a)\operatorname{Re}\{v(x,T)\}.$

This gives terminal conditions at $t = T$ so that Equations (3.9) may then be solved backward in time.

Next, setting functional derivatives with respect to the control variables to zero gives

(3.12a) $\quad \delta_\kappa \mathcal{J} = \int_0^T \operatorname{Re}\left\{\lambda^\dagger v + \mu^\dagger u\right\} dt - \gamma \partial_x^2 \kappa = 0,$

(3.12b) $\quad \delta_\eta \mathcal{J} = \int_0^T \operatorname{Re}\left\{\lambda^\dagger u + \mu^\dagger v\right\} dt - \gamma \partial_x^2 \eta = 0.$

Equations (3.12), together with the boundary conditions implied by the admissible class $\mathcal{C}$, give two-point boundary value problems over the domain $[0, a]$. Lastly, the vanishing of functional derivatives with respect to the costate variables $\lambda^\dagger$ and $\mu^\dagger$ returns the state equations (2.4), i.e., the NLCME.

We use a second order in time operator splitting method to solve the state equations (2.4) and to solve costate equations (3.9) backward from their terminal condition (3.11). The details of the method, which itself is a refinement on the splitting method used in [27], are given in Appendix A.3. Equations (3.12) are used in the computation of control gradients as required by the computational optimization method in Section A.2.

**3.3. Numerical Optimization Strategy.** In order to solve Problem (3.3), we use a *hybrid* optimization method; a combination of a global, non-convex method followed by a local, iterative method. The methodology we use in this paper is similar to one used by Sørensen, et al. [32], and allows for the use of a global search routine based on stochastic optimization to overcome non-convexity. Non-convex objective functions may, of course, possess many local minima, and a global method seeks to efficiently search for a near-optimal one. By then feeding results from the global method into the local one, convergence near the local minimum is accelerated.

The first step in the global search is to use a Chopped Random Basis (CRAB) method [9, 10] which efficiently reduces the numerical optimization problem to a nonlinear programming problem (NLP). This NLP is then solved via a genetic algorithm called Differential Evolution (DE) [33]. Implementation details of this method are given in Appendix A.1. The next step is to refine the result of CRAB/DE using a projected gradient descent due to von Winckel and Borzi [7], which we explain in detail in Appendix A.2.

**4. Optimization Results.** We now present the results of using the hybrid numerical optimization strategy on optimal control problem (3.3). In most of the results presented, we slightly change the optimization problem. We keep the raised-cosine portion of the profile fixed on $[0, L_1]$ and only optimize the second section of the profile on $[L_1, L_2]$. This is based on our initial experimentation that found optimization over the first segment made less difference. In all simulations, we find the Tikhonoff parameter $\gamma$ on the order of $10^{-6}$ to be satisfactory. All other parameters regarding numerical simulations are consistent with those used so far throughout this work. In our first simulation, we do not allow a chirp (fixing $\eta = 0$). Our hybrid method finds an apodization function which couples 74.1% of the incident light into the grating. This is shown in Figure 4.1. Because of the stochastic nature of the method, running the simulation again would find a different profile, but we found that if the parameters of the search were held fixed, the resulting efficiency was fairly stable.

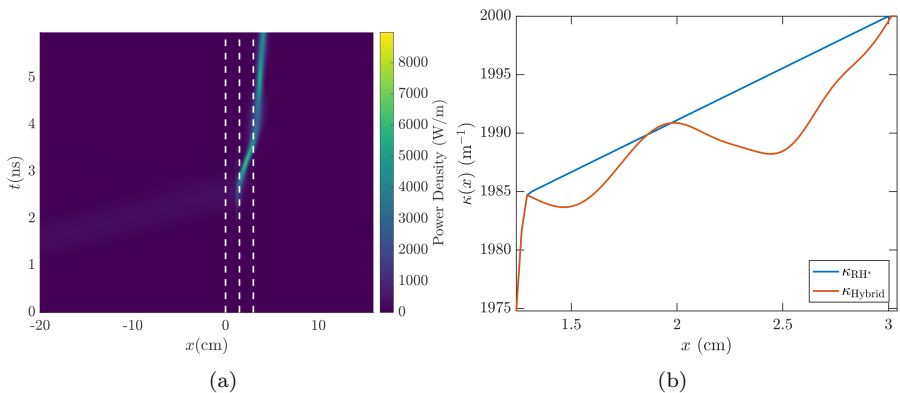

(a)      (b)

Fig. 4.1. *The result of using the hybrid method to find more efficient apodization functions $\kappa$ which are nearby the $(\xi, \zeta)_*$ apodization. 74.1% of the incident light is now coupled into the grating.*

Next, we include a chirp on the interval $[L_1, L_2]$(to be consistent with above), and show the computed optimizer in Figure 4.2: it couples 77.7% of the incident light. From the optimal apodization functions, shown in Figures 4.1–4.2, we observe that the nearby efficient apodization functions have a somewhat large negative gradient to the right of the optimization boundary point $x_0$. This hints at how more of the total apodization region should be allocated toward the shallow, adiabatic portion of the grating.

For this reason, we slightly relax the design constraint of a three centimeter apodization region, extend the width of the optimization domain to $[1.19, 3.4]$ cm, and perform the search again. The extension to the left is motivated by the above-mentioned observation on the nearby optimal apodization functions, and the extension to the right is ad hoc. We extend the domain so that the entire grating structure is still reasonably within some technological constraint, i.e., within a margin of 15% the total size of the Rosenthal and Horowitz apodization region, and, moreover, maximize performance. We show, in Figure 4.3, grating functions which now successfully couple 82.6% of the incident light. We also find that the resulting in-gap soliton has a group velocity with a magnitude 0.53% that of light speed. We emphasize this unintended, yet fortuitous improvement in the slowdown of the coupled light against the result of the original $(\xi, \zeta)_{\text{RH}}$ apodization visually in Figure 4.4.

The final result we show, in Figure 4.5, is found from a search on the entire optimization domain $[0, 3]$ cm. After several hours of computation time, the best grating structure found is 81.4% efficient. Moreover, this is found by looking for apodization profiles similar in feature to the designs shown in Figure 4.2. That is, we begin the search by looking for CRAB amplitudes that are each within 10% of the coefficients of a sine series decomposition of the grating structure in Figure 4.2. We believe this result clearly shows the value of Rosenthal and Horowitz's intuition in their design choices, since the best grating structure found from a general search over the entire domain $[0, 3]$ cm, including a chirp, was only about 68% efficient.

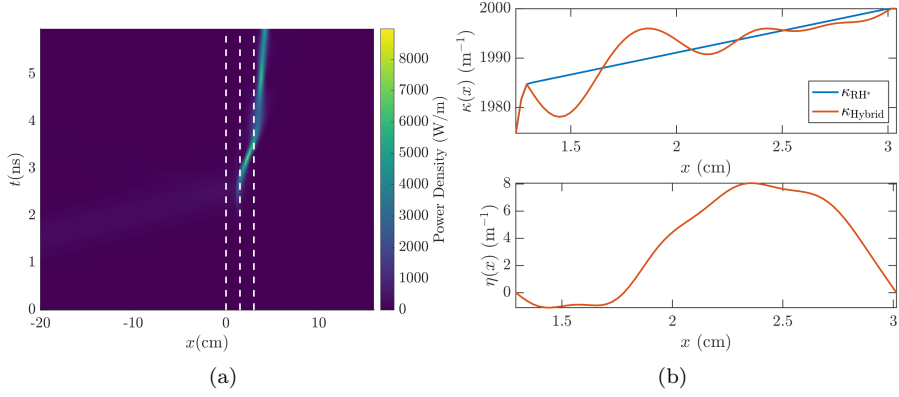

FIG. 4.2. *The result of including a chirp in the search for efficient grating functions near the $(\xi, \zeta)_*$ apodization. 77.7% of the incident light is now coupled into the grating.*

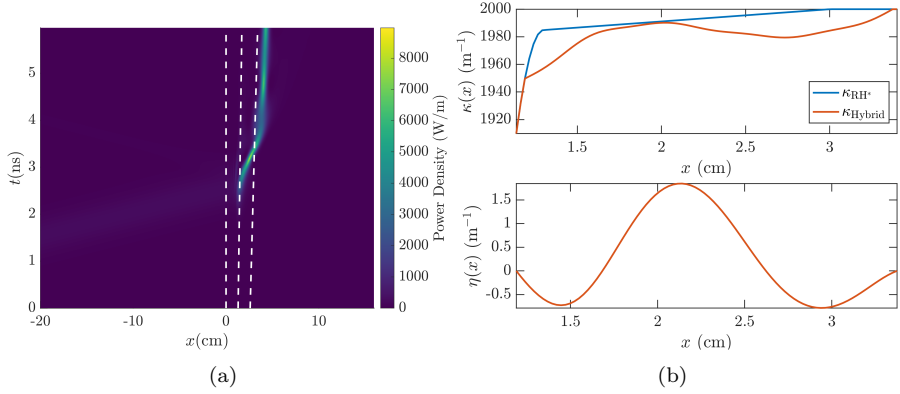

FIG. 4.3. *The result of relaxing the optimization domain to be slightly wider, i.e., $x_0 = 1.19$ cm and $a = 3.4$ cm. The FBG couples 82.6% of the light with a group velocity 0.53% the speed of light.*

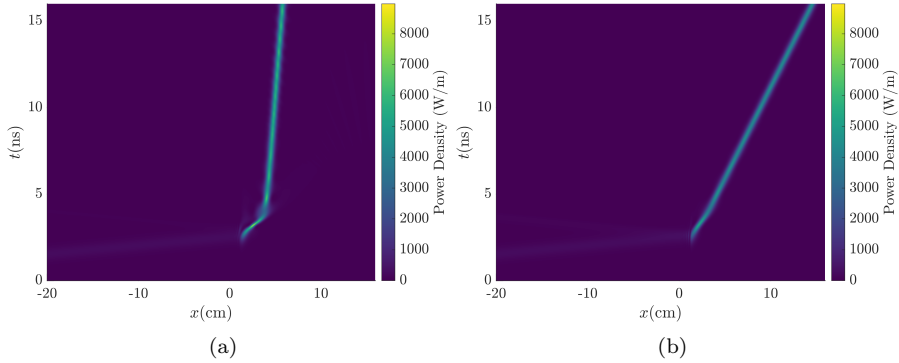

FIG. 4.4. *A long time simulation consistent with the results of Figure 4.3 emphasizing the dramatic reduction in the group velocity of the solitary wave.* **(a)** *The locally optimal grating structure shown in Figure 4.2,* **(b)** *The $(\xi, \zeta)_{\mathrm{RH}}$ apodization from Figure 2.2.*

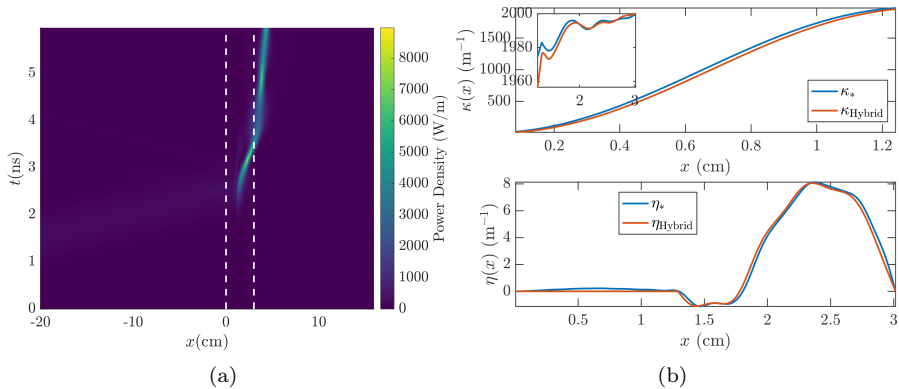

FIG. 4.5. *A result of using the hybrid method optimizing over the entire domain $x \in [0, 3]$cm. The coupling efficiency of the grating is about 81.4%. The notation $\kappa_*$ and $\eta_*$ denote the designs shown in Figure 4.2.*

**5. Concluding Remarks.** In this work, we formulate a simple and physically motivated optimal control problem aimed at efficiently coupling light into an FBG. By employing numerical optimization methods, widely used for example in the quantum control literature [10, 19, 32], we demonstrate the viability of optimal control theory in the design of FBGs which act as efficient compressors and pulse-delayers.

We optimize a previously reported design, and provide guidance on how to further explore the design space. In terms of practical application, we believe work due to Sahin, et al. [30, 31], demonstrates that the length scales of our designs are well within the range of modern fabrication methods which operate on the order of microns. Additionally, the methodology used here can be applied, with suitable modifications, to other problems constrained by dispersive equations, see for example our previous work on the manipulation of Bose-Einstein condensates and optical beam reshaping [3–5].

Considering the chirp of the grating as part of the design increases both the transmission of the light and the effectiveness of the grating as a pulse-delayer. Moreover, all of the best apodization functions we found are, loosely speaking, near the Rosenthal and Horowitz design. Since the admissible class $\mathcal{C}$ is so large, we cannot claim these results are globally optimal despite the extensive search we performed. We also observe that in cases where we see significant gains in the transmission of light into the fiber, we find this comes at the cost of decoherence through radiation buildup.

This radiation buildup is most clearly seen after performing a fit of the Bragg soliton formula (2.7) to the simulation data from Figure 4.3. We observe significant oscillatory tails in the real and imaginary parts of the solitary wave in Figure 5.1. We find that the grating structure is only 78.1% effective in terms of *coherent* energy since 94.5% of the total transmitted energy belongs to the Bragg soliton rendering this result to be less impressive.

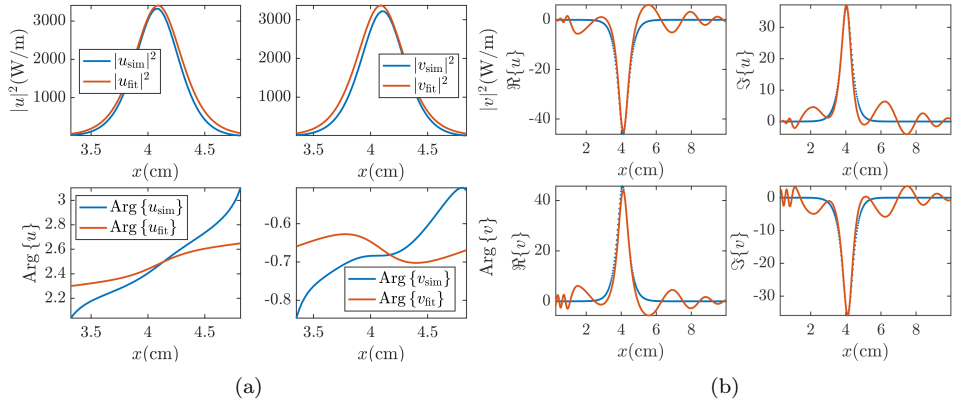

FIG. 5.1. *A Bragg soliton fitting with (dimensionless) parameters $c = 0.0058$, $\theta = 0.1806$ corresponding to the optimal grating structures in Figure 4.3 at $t = 4.8$ns. **(a)**: Shown in power density and local phase. **(b)**: Shown in real and imaginary parts. Evidence of solitary wave decoherence is seen to be present.*

Secondly, it is noted that the discovery of a scenario which produces very slow light is promising for applications. However, this is an unintended byproduct of maximizing the flux of energy. One future direction of study is to use optimal control to explicitly engineer for the combination of slow light and high energy. We believe this can be done by a simultaneous maximization of the energy flux and minimization of the momentum flux corresponding to Equation (2.10). Such a study is planned for future work.

We further believe our numerical results, in particular those of Figure 4.3, serve as an impetus for investigating the coupled-mode dynamics more thoroughly using soliton-specific methods. In one direction, there is mathematical work to be done to understand the adiabatic interaction between the grating and the excited waveform using soliton perturbation theory. In another, optimization methods based on nearby integrable systems, such as a nonlinear Schrödinger or Thirring model, may be conceived as ways to filter, through the spectrum of the associated Lax operators, the radiation which persists in our numerical simulations. We leave these thoughts as subject for future work.

**Acknowledgments.** We would like to thank Alejandro Aceves, Andrea Blanca-Redondo, Boris Malomed, Richard Moore, Stephen Shipman, David Shirokoff, and Michael Weinstein, for very helpful discussions about our work and preparing the resulting manuscript for publication.

**Appendix A. Numerical Methods.**

**A.1. Global Optimization Method.** In the first step of the hybrid method, we represent the apodization profile using a Galerkin approximation. This results in a so-called Chopped Random Basis (CRAB) method which reduces the complexity of the optimal control problem so that standard non-convex nonlinear programming (NLP) techniques can be applied. It relies on choosing controls from the span of an appropriately chosen finite set of basis functions so that the optimization is performed over a relatively small set of unknown coefficients. The basis is chosen so that controls remain in the appropriate admissible space $\mathcal{C}$ in the context of the control problem (3.3).

A common representation in the CRAB method is of the form

$$\beta_r(x) = \mathcal{P}(x; \beta(x_0), \beta(a), x_0, a) + \sum_{j=0}^{N-1} \varepsilon_j \varphi_j(x; x_0, a), \quad x \in [x_0, a]. \tag{A.1}$$

Here, $\beta$ denotes either grating function $\kappa$ or $\eta$, $\mathcal{P}$ is a fixed function satisfying the boundary conditions of the admissible class (3.2), each $\varphi_j(x)$ is a basis function with vanishing boundary conditions, and the coefficients $\varepsilon_j$ are parameters to be optimized over. It is clear that if the polynomial $\mathcal{P}$ and the basis functions $\varphi_j$ are chosen well enough, then control ansatz (A.1) reliably simplifies the optimal control problem.

More specifically, the CRAB ansatz we use throughout this work is

(A.2a)
$$\kappa_r(x) = (\kappa_0 - \kappa_{\text{RH}}(x_0)) \sum_{j=1}^{N} \frac{r_j^\kappa}{j^2} \sin\left(j\pi \frac{x-x_0}{a-x_0}\right) + (\kappa_0 - \kappa_{\text{RH}}(x_0)) \frac{x-x_0}{a-x_0} + \kappa_{\text{RH}}(x_0),$$

$$\eta_r(x) = \frac{\kappa_0}{100} \sum_{j=1}^{N} \frac{r_j^\eta}{j^2} \sin\left(j\pi \frac{x-x_0}{a-x_0}\right), \tag{A.2b}$$

on the optimization domain $x \in [x_0, a]$, where each $r_j^\kappa$ and $r_j^\eta$ is a random variable drawn uniformly from $[-1, 1]$, $\kappa_{\text{RH}}$ is a Rosenthal and Horowitz apodization (2.11) with free parameters $(\xi, \zeta)$, and $\kappa_0 = 2\text{mm}^{-1}$, consistent with the constant grating portion of $\kappa_{\text{RH}}$. From experience with the numerical optimization, we find that chirp functions which are about two orders of magnitude smaller than $\kappa_0$ perform well; this is why we include the factor of 100 in the CRAB ansatz for $\eta$. A modest number of basis functions, $N = 15$, has proven adequate here.

To solve the resulting NLP problem, we use differential evolution (DE) [33]. DE is a stochastic optimization method used to search for candidate solutions to non-convex optimization problems. DE is a so-called genetic algorithm that draws inspiration from evolutionary genetics. DE searches the space of candidate solutions by initializing a population set of vectors, known as agents, within some chosen region of the search space. These vectors are then randomly mutated into a new population set, or generation. The mutation operates via two mechanisms: a weighted combination, with parameter $p_{\text{weight}} \in (0, 2)$, and a "crossover," with parameter $p_{\text{cross}}$ which randomly exchanges "traits", or elements, between agents. We find a population size of 50 agents, a weight $p_{\text{weight}} = 0.8$, and crossover $p_{\text{cross}} = 0.9$ to be effective.

DE ensures that the objective functional $\mathcal{J}$ decreases monotonically with each generation. As each iteration "evolves" into the next, inferior vectors "inherit" optimal traits from superior vectors via mutations. DE only allows mutations which are more optimal with respect to $\mathcal{J}$ to pass to the next generation. After a sufficient number of iterations, the best vector in the final generation is chosen as the candidate solution most likely to be globally optimal with respect to an objective functional. We find 30 iterations to be sufficient in this regard.

Genetic algorithms, which require very few assumptions about the objective functional, are part of a wider class of optimization methods called metaheuristics. Although metaheuristics are useful for non-convex optimization problems, they do not make guarantees about the global optimality of candidate solutions. Since the algorithm is stopped after a finite number of iterations, different random realizations return different candidate optimizers. For this reason, we use DE to search for candidate solutions and use these candidates in order to generate initial controls, through the CRAB representation (A.1), for an iterative method which guarantees local optimality up to some threshold. We note that in practice, the best of five or so realizations through the CRAB/DE method is sufficient before moving on to the refinement stage.

**A.2. Local Optimization Method.** We use a line search strategy due to von Winckel and Borzi [36]. The von Winckel and Borzi (vWB) method is an appropriate generalization of the well-known gradient descent method from $\mathbb{R}^n$ to an appropriate affine function space which automatically preserves the boundary conditions of the admissible class $\mathcal{C}$ mentioned in the context of optimal control problem (3.3). This method has been frequently applied in the quantum control literature; see for example [19, 23, 32].

For ease of notation, we describe the vWB method for optimizing the apodization function $\kappa$, since its extension to $\eta$ is trivial. Recall the optimal control problem we want to solve is, in unconstrained form, given by Equation (3.7). The method of gradient descent, in this context, is given by following iteration

$$\text{(A.3)} \qquad \kappa_{k+1} = \kappa_k - \alpha_k \nabla_\kappa \mathcal{L} \big|_{\kappa = \kappa_k},$$

where the linear operator $\nabla_\kappa$ is the gradient, or Fréchet derivative, of the Lagrangian $\mathcal{L}$ with respect to the control $u$. The stepsize $\alpha_k$ is chosen adaptively via the Armijo-Goldstein condition [8].

Recall that the definition of a Fréchet derivative depends on the choice of function space in which it is to be understood. If the Fréchet derivative is understood in the sense of $L^2([x_0, a])$, then it can be identified with the functional derivative of the objective $\mathcal{J}$, which in this case can be shown to be

$$\text{(A.4)} \qquad \delta_\kappa \mathcal{J} = \int_0^T \text{Re}\left\{\lambda^\dagger v + \mu^\dagger u\right\} dt - \gamma \partial_x^2 \kappa.$$

Note that this coincides with the Euler-Lagrange equation $\delta_\kappa \mathcal{J} = 0$ given by Equation (3.12). If this choice is made, however, the increment $\alpha_k \nabla_\kappa \mathcal{L}\big|_{\kappa=\kappa_k}$ would not in general satisfy the boundary conditions on the control $\kappa_k$, and the updated control $\kappa_{k+1}$ would leave the admissible set $\mathcal{C}$. This problem is avoided by using a different function space $X$ to define the operator $\nabla_\kappa$.

To this end, consider an arbitrary displacement $\nu \in C_c^\infty([x_0, a])$ and an arbitrary $\varepsilon > 0$. We know Taylor's theorem holds, i.e., the series

$$\tag{A.5} \mathcal{J}[\kappa + \varepsilon\nu] = \mathcal{J}[\kappa] + \varepsilon \langle \nabla_\kappa \mathcal{L}(\kappa), \nu \rangle_X + \mathcal{O}(\varepsilon^2)$$

holds term-by-term independently of the choice of the Hilbert space $X$ for sufficiently regular functionals $\mathcal{J}$. The vWB method chooses the homogeneous and traceless Sobolev space $\dot{H}_0^1([x_0, a])$ for $X$. By equating the directional, or Gateaux, derivatives with respect to $L^2([x_0, a])$ and with respect to $\dot{H}_0^1([x_0, a])$, we see that

$$\tag{A.6} \begin{aligned} \langle \nabla_\kappa \mathcal{L}, \nu \rangle_{L^2([x_0,a])} &= \langle \delta_\kappa \mathcal{J}, \nu \rangle_{L^2([x_0,a])} \\ &= \langle \nabla_\kappa \mathcal{L}, \nu \rangle_{\dot{H}_0^1([x_0,a])} := \int_{x_0}^a \partial_x \nabla_\kappa \mathcal{L} \partial_x \nu \, dx = -\langle \partial_x^2 \nabla_\kappa \mathcal{L}, \nu \rangle_{L^2([x_0,a])}, \end{aligned}$$

where an integration by parts is used once along with the boundary conditions on $\nu$.

Since this holds for all displacements $\nu \in C_c^\infty([x_0, a])$, we conclude, by the fundamental lemma of the calculus of variations [15], the strong form of Equation (A.6)

$$\tag{A.7} -\partial_x^2 \nabla_\kappa \mathcal{L} = \delta_\kappa \mathcal{J}, \quad \nabla_\kappa \mathcal{L}(x_0) = \nabla_\kappa \mathcal{L}(a) = 0,$$

also holds. This renders an admissible gradient whose homogeneous Dirichlet conditions are induced by choosing increments specifically from the traceless space $\dot{H}_0^1([x_0, a])$. In order to solve the boundary value problem (A.7) for the control gradient $\nabla_\kappa \mathcal{L}$, we use Chebyshev collocation [35].

**A.3. Solvers for the Evolution Equations.** Here we show how we numerically solve the NLCME (2.4), and costate equations (3.9). We follow an operator splitting method originally due to Rosenthal and Horowitz [27], and improve the order of convergence in time to second order. This improvement is critical in an optimization since it cuts the number of necessary spatial discretization points in half, which also effectively cuts the computation time of evaluating the functional (3.3) in half; searches for optimal grating functions is on the order of hours. Note that the splitting methods of this section fundamentally assume the solutions of Equations (2.4) are analytic. Although this is a strong assumption to make given that the grating functions are only assumed to be absolutely continuous, we believe the resulting numerical simulations justify this assumption a posteriori.

One of the advantages of operator splitting methods is in the ease of implementation, especially if each operator has an explicit update. Most operator splitting methods split the linear and nonlinear operators into two separate operators. However, since the numerical model, discussed in Section 2, requires signalling data without a backpropagating wave $v(x, t)$, this means solving the state equations (2.4) using a method which splits the linear and nonlinear operators is not possible; $v$ is excited by the coupling through the linear *and* nonlinear terms.

Instead, we follow the method of Rosenthal and Horowitz which splits Equation (2.4) into

$$\tag{A.8} \partial_t \begin{pmatrix} u \\ v \end{pmatrix} = (A + B) \begin{pmatrix} u \\ v \end{pmatrix},$$

where the matrix operator $A$ is defined as

$$\tag{A.9} A = i \begin{pmatrix} -\partial_x \bullet + \left(|\bullet|^2 + 2|v|^2\right) \bullet & 0 \\ 0 & \partial_x \bullet + \left(2|u|^2 + |\bullet|^2\right) \bullet \end{pmatrix},$$

using • to help express how the operator $A$ functions, and where the matrix $B$ is

$$B = i \begin{pmatrix} \eta & \kappa \\ \kappa & \eta \end{pmatrix}. \tag{A.10}$$

Computations involving the matrix exponential of $B$ are simple. Using the Taylor series definition of the matrix exponential and diagonalizing the operator $B$ into canonical form, we have

$$\begin{aligned} e^{B\Delta t} &= \sum_{n=0}^{\infty} \frac{(B\Delta t)^j}{j!} = P \sum_{n=0}^{\infty} \frac{(\Delta t \Lambda)^n}{n!} P^{-1} = P e^{i\Delta t \Lambda} P^{-1} \\ &= e^{i\Delta t \eta} \begin{pmatrix} \cos(\kappa \Delta t) & i\sin(\kappa \Delta t) \\ i\sin(\kappa \Delta t) & \cos(\kappa \Delta t) \end{pmatrix}, \end{aligned} \tag{A.11}$$

where the matrix $\Lambda$ is the $2 \times 2$ diagonal eigenvalue matrix of $B$, and $P$ is a $2 \times 2$ matrix whose columns are eigenvectors of $B$.

The computation of $e^{A\Delta t}$ explicitly assumes analyticity of the coupled modes $u$ and $v$. Observe the effect of the following operator on monomials:

$$e^{a\partial_x} x^m = \sum_{k=0}^{\infty} \frac{(a\partial_x)^k}{k!} x^m = \sum_{k=0}^{m} \binom{m}{k} a^k x^{m-k} = (x+a)^m. \tag{A.12}$$

The effect of this operator is simply a spatial translation of $a$ units. Therefore, it is easy to show the effect of this operator on a function $f(x)$, analytic in a neighborhood about $x = 0$, is

$$e^{a\partial_x} f(x) = f(x+a). \tag{A.13}$$

This implies directly that

$$e^{A\Delta t} \begin{pmatrix} u(x,t) \\ v(x,t) \end{pmatrix} = \begin{pmatrix} e^{i\Delta(|u(x-\Delta t,t)|^2 + 2|v(x,t)|^2)} u(x-\Delta t, t) \\ e^{i\Delta(2|u(x,t)|^2 + |v(x+\Delta t,t)|^2)} v(x+\Delta t, t) \end{pmatrix} \tag{A.14}$$

is simply an appropriately coupled, advective update.

Clearly, this method also couples the spatial and temporal discretizations, and for this reason, we follow Rosenthal and Horowitz in choosing $\Delta t = \Delta x := \Delta$. The first-order composition of propagators, after $M$ time steps, is straightforward:

$$e^{(A+B)M\Delta} = e^{B\Delta} e^{A\Delta} \ldots e^{B\Delta} e^{A\Delta} + \mathcal{O}(\Delta). \tag{A.15}$$

The improvement we make on this method is facilitated by using the symmetric average suggested by MacNamara and Strang [22]:

$$e^{(A+B)M\Delta} = \frac{1}{2^M} \left( e^{A\Delta} e^{B\Delta} + e^{B\Delta} e^{A\Delta} \right) \ldots \left( e^{A\Delta} e^{B\Delta} + e^{B\Delta} e^{A\Delta} \right) + \mathcal{O}(\Delta^2), \tag{A.16}$$

where the matrix exponential $e^{A\Delta} e^{B\Delta}$ is computed by similar means as computing $e^{B\Delta} e^{A\Delta}$. Note that the often used Strang splitting, after taking $M$ steps,

$$e^{(A+B)M\Delta} = e^{A\Delta/2} e^{B\Delta} e^{A\Delta} \ldots e^{A\Delta} e^{B\Delta} e^{A\Delta/2} + \mathcal{O}(\Delta^2), \tag{A.17}$$

is not easily available because of the coupling of discretizations $\Delta t$ and $\Delta x$, otherwise, about half of the computational effort could be saved.

As required by the vWB method of Section A.2, we must also solve the costate, or coupled-mode adjoint, equations

$$i\partial_t \lambda + i\partial_x \lambda + \left(\eta + 2\mathcal{E} + u^{\dagger 2}\right)\lambda + \left(\kappa + 4v^\dagger \operatorname{Re}\{u\}\right)\mu = 0, \quad \text{(A.18a)}$$
$$i\partial_t \mu - i\partial_x \mu + \left(\eta + 2\mathcal{E} + v^{\dagger 2}\right)\mu + \left(\kappa + 4u^\dagger \operatorname{Re}\{v\}\right)\lambda = 0, \quad \text{(A.18b)}$$
$$\lambda(x,T) = 2iH(x-a)\operatorname{Re}\{u(x,T)\}, \quad \text{(A.18c)}$$
$$\mu(x,T) = 2iH(x-a)\operatorname{Re}\{v(x,T)\}, \quad \text{(A.18d)}$$

derived in Section 3. To this end, we use the splitting

$$\partial_t \begin{pmatrix} \lambda \\ \mu \end{pmatrix} = (C + D) \begin{pmatrix} \lambda \\ \mu \end{pmatrix}, \quad \text{(A.19)}$$

where the matrix operators $C$ and $D$ are given by

$$C = \begin{pmatrix} -\partial_x & 0 \\ 0 & \partial_x \end{pmatrix}, \qquad D = i \begin{pmatrix} d_{11} & d_{12} \\ d_{21} & d_{22} \end{pmatrix} \quad \text{(A.20)}$$

where

$$d_{11}(x,t) = \eta + 2\left(|u|^2 + |v|^2\right) + u^{\dagger 2},$$
$$d_{12}(x,t) = \kappa + 4v^\dagger \operatorname{Re}\{u\},$$
$$d_{21}(x,t) = \kappa + 4u^\dagger \operatorname{Re}\{v\},$$
$$d_{22}(x,t) = \eta + 2\left(|u|^2 + |v|^2\right) + v^{\dagger 2}.$$

In order to evolve the costates, we must do so backward in time, since the terminal conditions (A.18c)–(A.18d) are specified at the final time of simulation. The first-order backward composition of operators

$$e^{-(C+D)M\Delta} = e^{-D\Delta}e^{-C\Delta}\ldots e^{-D\Delta}e^{-C\Delta} + \mathcal{O}(\Delta), \quad \text{(A.21)}$$

is straightforward to form through computations of

$$e^{-C\Delta}\begin{pmatrix}\lambda(x,t)\\ \mu(x,t)\end{pmatrix} = \begin{pmatrix}\lambda(x+\Delta,t)\\ \mu(x-\Delta,t)\end{pmatrix} \quad \text{(A.22)}$$

and
(A.23)
$$e^{-D\Delta} = e^{-\frac{i\Delta}{2}(d_{11}+d_{22})}\begin{pmatrix}\cos\left(\frac{\varphi\Delta}{2}\right) - \frac{i\phi}{\varphi}\sin\left(\frac{\varphi\Delta}{2}\right) & -\frac{2id_{12}}{\varphi}\sin\left(\frac{\varphi\Delta}{2}\right) \\ -\frac{2id_{21}}{\varphi}\sin\left(\frac{\varphi\Delta}{2}\right) & \cos\left(\frac{\varphi\Delta}{2}\right) + \frac{i\phi}{\varphi}\sin\left(\frac{\varphi\Delta}{2}\right)\end{pmatrix},$$

where

$$\phi = d_{11} - d_{22}, \qquad \varphi = \sqrt{\phi^2 + 4d_{12}d_{21}}.$$

Using the symmetric average (A.16) as before, and computing $e^{-D\Delta}e^{-C\Delta}$ via similar means, yields a second-order in time method for the adjoint coupled-mode equations (A.18).